\documentclass[prl,aps,twocolumn,showpacs]{revtex4-1}
\usepackage{amsfonts}
\usepackage{amsmath}
\usepackage{amssymb}
\usepackage{graphicx}%
\providecommand{\U}[1]{\protect\rule{.1in}{.1in}}
\providecommand{\U}[1]{\protect\rule{.1in}{.1in}}

\begin{document}
\preprint{ }
\title{Singlet and Triplet Superfluid Competition in a Mixture of Two-Component Fermi and One-Component Dipolar Bose gases}
\author{Ben Kain and Hong Y. Ling}
\affiliation{Department of Physics and Astronomy, Rowan University, Glassboro, New Jersey 08028}

\begin{abstract}
We consider a mixture of two-component Fermi and (one-component) dipolar Bose
gases in which both dipolar interaction and s-wave scattering between fermions
of opposite spins are tunable. We show that in the long wavelength limit, the
anisotropy in the Fermi-Fermi interaction induced by phonons of the dipolar
condensate can strongly enhance the scattering in the triplet channel. \ We
investigate in detail the conditions for achieving optimal critical
temperature at which the triplet superfluid begins to compete with the singlet superfluid.

\end{abstract}
\maketitle

The ability to easily mix cold atoms of different species to form new quantum
systems brings another exciting dimension to the study of ultracold atomic
gases. \ A single-component Fermi gas only supports Cooper pairing with odd
parities, such as p-wave pairing, which are typically strongly suppressed in
accordance with Wigner's threshold law \cite{landau89}. Mixing bosons induces
an attractive interaction between fermions \cite{bijlsma00}, which, as Efremov
and Viverit \cite{efremov02} pointed out, raises the prospect of achieving
p-wave superfluidity in a Fermi-Bose (FB) mixture. Recently, Dutta and
Lewenstein \cite{dutta10} generalized the idea to a (2D) mixture involving
dipolar bosons with the goal of realizing a superfluid with $p_{x}+ip_{y}$
symmetry whose excitations are non-Abelian anyons that are the building blocks
for topological quantum computation \cite{nayak08}, and Nishida
\cite{nishida09} sought the same goal by mixing fermion gases of different
species in different dimensions.

In this Letter, we consider a (3D) homogeneous mixture (with an effective
volume $V$) between a \textit{two-component }Fermi gas and a \textit{dipolar}
Bose gas, made up of two equally populated (balanced) hyperfine spin states
($\left\vert \uparrow\right\rangle $ and $\left\vert \downarrow\right\rangle
$) of a non-dipolar fermionic atom of mass $m_{F}\,$, and a ground state of a
bosonic molecule (or atom) of mass $m_{B}$ with an induced dipole aligned
along the external electric field direction $z$. \ The two pseudo spins
provide fermions with the opportunity to pair not only\ via triplet (with odd
parities) but also via singlet (with even parities) channels of interaction.
This opens up the possibility of using this two-component FB model to emulate
and explore pairing physics analogous to that in superfluid $^{3}%
$He\ \cite{vollhardt90}, which is known to be greatly enriched by the
existence of an internal spin degree of freedom.

In the context of ultracold atomic physics, there has been a recent upsurge of
activity in pursuing similar goals but with\ (3D) two-component dipolar Fermi
gases \cite{samokhin06,shi10}, motivated largely by recent rapid experimental
advancement in achieving ultracold dipolar gases both in $^{40}$K -$^{87}$Rb
polar molecules \cite{ospelkaus08}, and in Cr \cite{stuhler05} and spin-1 Rb
atoms \cite{vengalattore08}. \ Such studies \cite{samokhin06,shi10} represent
a generalization of earlier work \cite{marinescu98,baranov02} aimed at
exploiting a $d_{r^{2}-3z^{2}\text{ }}$-type of anisotropy in dipole-dipole
interactions for enhancing triplet pairing in single-component dipolar Fermi systems.

The induced Fermi-Fermi interaction mediated by a dipolar condensate is also
anisotropic in nature and thus opens up a new avenue for studying superfluids
with unusual pairings.\ The progress in this area has so far been limited, to
the best of our knowledge, to a single-component model in a 2D geometric
setting \cite{dutta10}. \ In contrast, the present work expands such studies
to a 3D two-component mixture, where both the dipolar interaction between
bosons and the s-wave scattering between fermions of opposite spins are
independently tunable, and seeks to use it as a model to explore the physics
that are currently being hotly pursued in two-component dipolar Fermi gas
systems \cite{shi10}. \ In this Letter, we study in detail the anisotropic
nature of the 3D induced interaction, and particularly how one should prepare
a two-component FB mixture in order to maximize the opportunity this induced
interaction affords for raising critical temperatures at which phases of
different parities begin to compete.

To begin with, we model our system with a (grand canonical) Hamiltonian
$\hat{H}=\hat{H}_{B}+\hat{H}_{BF}+\hat{H}_{F}$:
\begin{align}
\hat{H}_{B}    &=\sum_{\mathbf{k}}\left(  \xi_{\mathbf{k},B}\equiv
\epsilon_{\mathbf{k},B}-\mu_{B}\right)  \hat{b}_{\mathbf{k}}^{\dag}\hat
{b}_{\mathbf{k}}+\left(  2V\right)  ^{-1}\times\nonumber\\
&\quad  \sum_{\mathbf{k},\mathbf{k}^{\prime},\mathbf{q}}\left[  U_{BB}%
+U_{DD}\left(  \mathbf{q}\right)  \right]  \hat{b}_{\mathbf{k}+\mathbf{q}%
}^{\dag}\hat{b}_{\mathbf{k}^{\prime}-\mathbf{q}}^{\dag}\hat{b}_{\mathbf{k}%
^{\prime}}\hat{b}_{\mathbf{k}},\label{HB}\\
\hat{H}_{BF}    &=U_{BF}\left(  2V\right)  ^{-1}\sum_{\mathbf{k}%
,\mathbf{k}^{\prime},\mathbf{q},\sigma}\hat{a}_{\sigma,\mathbf{k}}^{\dag}%
\hat{a}_{\sigma,\mathbf{k}+\mathbf{q}}\hat{b}_{\mathbf{k}^{\prime}}^{\dag}%
\hat{b}_{\mathbf{k}^{\prime}-\mathbf{q}},\label{HI}\\
\hat{H}_{F}    &=  \sum_{\mathbf{k},\sigma}\left(  \xi_{\mathbf{k},F}%
\equiv\epsilon_{\mathbf{k},F}-\mu_{F}\right)  \hat{a}_{\mathbf{k},\sigma
}^{\dag}\hat{a}_{\mathbf{k},\sigma}+\left(  2V\right)  ^{-1}\times\nonumber\\
&  \sum_{\mathbf{k},\mathbf{k}^{\prime},\mathbf{q,\sigma,\sigma}^{\prime}%
}U_{\sigma\sigma^{\prime}}\left(  \mathbf{q}\right)  \hat{a}_{\mathbf{k}%
+\mathbf{q,\sigma}}^{\dag}\hat{a}_{\mathbf{k}^{\prime}-\mathbf{q}%
,\sigma^{\prime}}^{\dag}\hat{a}_{\mathbf{k}^{\prime},\sigma^{\prime}}\hat
{a}_{\mathbf{k},\sigma}, \label{HFF}%
\end{align}
where $\hat{b}_{\mathbf{k}}$ ($\hat{a}_{\sigma,\mathbf{k}}$) is the field
operator for annihilating a boson (a fermion of spin $\sigma)$ of kinetic
energy $\epsilon_{\mathbf{k},B}=\hbar^{2}k^{2}/2m_{B}$ $\left(  \epsilon
_{\mathbf{k},F}=\hbar^{2}k^{2}/2m_{F}\right)  $ and chemical potential
$\mu_{B}$ ($\mu_{F}$). The low temperature physics of the mixture under
consideration arises from the interplay between short- and long-range two-body
interactions. \ The former is dominated by s-wave scattering characterized
with strengths: $U_{BB}=4\pi\hbar^{2}a_{BB}/m_{B}$, $U_{FF}=4\pi\hbar
^{2}a_{FF}/m_{F}$, $U_{BF}=4\pi\hbar^{2}a_{BF}/m_{BF}[\equiv2m_{B}m_{F}%
/(m_{B}+m_{F})]$, and $U_{\sigma\sigma^{\prime}}\left(  \mathbf{q}\right)
=U_{FF}\delta_{\sigma^{\prime},-\sigma}$, where $a_{BB},$ $a_{BF}$ and
$a_{FF}$ are the related scattering lengths, and the Kronecker-$\delta$
function in $U_{\sigma\sigma^{\prime}}\left(  \mathbf{q}\right)  $ limits the
fermionic s-wave interactions to fermions of opposite spins. \ The latter is
the dipole-dipole interaction (restricted to bosons) given by $U_{DD}\left(
\mathbf{q}\right)  =8\pi d^{2}P_{2}\left(  \cos\theta_{\mathbf{q}}\right)  /3$
in momentum space, with $d^{2}$ the dipolar interaction strength,
$P_{2}\left(  x\right)  =\left(  3x^{2}-1\right)  /2$ the second-order
Legendre polynomial, and $\theta_{\mathbf{q}}$ ($\phi_{\mathbf{q}}$) the polar
(azimuthal) angle of vector $\mathbf{q}$.

In the low temperature regime considered in the present Letter, bosons are
virtually all condensed to the zero-momentum mode and a straightforward
application of the Bogoliubov approximation, in which $\hat{b}_{\mathbf{k}=0}$
is treated as a c-number $b_{\mathbf{k}=0}$, yields a well-known picture of
the bosonic system described by Eq. (\ref{HB}): it consists of a collection of
phonon modes with the\ Bogoliubov dispersion relation $E_{\mathbf{k}}%
=v_{B}\hbar k\sqrt{1+\left(  \xi_{B}k\right)  ^{2}+2\varepsilon_{dd}%
P_{2}\left(  \cos\theta_{\mathbf{k}}\right)  }$ \cite{goral00} and a
homogeneous dipolar condensate with density $n_{B}=\left\vert b_{\mathbf{k}%
=0}\right\vert ^{2}$ ($\mu_{B}=n_{B}U_{BB}$), which is stable against collapse
provided $\varepsilon_{dd}$ $<$ $1$, above which phonons with $k\rightarrow0$
acquire imaginary frequencies. \ Here, $\varepsilon_{dd}=4\pi d^{2}/(3U_{BB})$
\cite{dell04} measures the strength of the dipolar interaction relative to the
s-wave interaction, $v_{B}=\sqrt{n_{B}U_{BB}/m_{B}}$ is the phonon speed, and
$\xi_{B}$ $=\hbar/\sqrt{4m_{B}n_{B}U_{BB}}$ is the healing length.
\ Integrating away the phonon degrees of freedom \cite{bijlsma00} leads to an
effective Fermi system described by the same Hamiltonian as Eq. (\ref{HFF}),
except that $U_{\sigma\sigma^{\prime}}\left(  \mathbf{k}\right)  =U_{FF}%
\delta_{\sigma^{\prime},-\sigma}+U_{ind}\left(  \mathbf{k}\right)  $, where
\begin{equation}
U_{ind}\left(  \mathbf{k}\right)  =-\frac{U_{BF}^{2}/U_{BB}}{1+\left(  \xi
_{B}k\right)  ^{2}+2\varepsilon_{dd}P_{2}\left(  \cos\theta_{\mathbf{k}%
}\right)  } \label{UIND}%
\end{equation}
is the phonon-induced Fermi-Fermi interaction in the static limit
\cite{bijlsma00,efremov02}. As can be seen, the induced interaction depends on
the dipole orientation differently than the direct dipole-dipole interaction
and therefore provides an alternative model for the exploration of spin
singlet and triplet paring.

Typical BCS mean-field theory proceeds with the introduction of the matrix
representation for the BCS order parameter in the uncoupled spin space:
$\Delta_{\sigma^{\prime}\sigma}\left(  \mathbf{k}\right)  =\sum_{\mathbf{k}%
^{\prime}}U_{\sigma\sigma^{\prime}}\left(  \mathbf{k}-\mathbf{k}^{\prime
}\right)  \left\langle \hat{a}_{-\mathbf{k}^{\prime},\sigma^{\prime}}\hat
{a}_{\mathbf{k}^{\prime},\sigma}\right\rangle $. \ The notion of spin singlet
and triplet pairing emerges when one moves from uncoupled to coupled spin
space spanned by a spin singlet $\left\vert S=0,M=0\right\rangle $ and triplet
$\left\vert S=1,M=-1,0,+1\right\rangle $, which are antisymmetric and
symmetric with respect to the spin exchange, respectively, where $M$ is the
$z$ projection of the total spin $S$. \ As one may easily verify, the gap
parameter, $\Delta^{s}\left(  \mathbf{k}\right)  =\left(  \Delta
_{\uparrow\downarrow}\left(  \mathbf{k}\right)  -\Delta_{\downarrow\uparrow
}\left(  \mathbf{k}\right)  \right)  /2$ associated with the singlet $\left(
S=0\right)  $ state is an even function of $\mathbf{k}$, while the three gap
parameters, $\Delta^{t,x}\left(  \mathbf{k}\right)  =\left(  \Delta
_{\downarrow\downarrow}\left(  \mathbf{k}\right)  -\Delta_{\uparrow\uparrow
}\left(  \mathbf{k}\right)  \right)  /2,$ $\Delta^{t,y}\left(  \mathbf{k}%
\right)  =\left(  \Delta_{\downarrow\downarrow}\left(  \mathbf{k}\right)
+\Delta_{\uparrow\uparrow}\left(  \mathbf{k}\right)  \right)  /2i$, and
$\Delta^{t,z}\left(  \mathbf{k}\right)  =\left(  \Delta_{\uparrow\downarrow
}\left(  \mathbf{k}\right)  +\Delta_{\downarrow\uparrow}\left(  \mathbf{k}%
\right)  \right)  /2$, associated with the triplet ($S=1$) states are odd
functions of $\mathbf{k}$, in accordance with Fermi statistics, where use of
$\Delta_{\alpha\beta}\left(  \mathbf{k}\right)  =\Delta^{s}\left(
\mathbf{k}\right)  i\left(  \sigma_{y}\right)  _{\alpha\beta}+\sum
_{u=x,y,z}\Delta^{t,u}\left(  \mathbf{k}\right)  i\left(  \sigma_{u}\sigma
_{y}\right)  _{\alpha\beta}$, a convention in the study of superfluid $^{3}$He
\cite{vollhardt90}, has been made, with $\sigma_{u}$ being the usual Pauli matrices.

To highlight the dominant physics, we ignore the Fermi surface deformation due
to the anisotropy of the Fermi-Fermi interaction \cite{miyakawa08} and
consider two-body scattering \ up to the level of the Born approximation
\cite{marinescu98,baranov02}, both of which hold in the weakly interacting
regime where $n_{F}U_{BF}^{2}/U_{BB}\ll$ $\epsilon_{F} = \text{Fermi energy}$. \ At
critical temperatures where the gaps are small, one can ignore the nonlinear
coupling between parings of different parities so that the critical
temperatures can be estimated by a set of linearly coupled gap equations
\cite{shi10,baranov02}: \
\begin{equation}
\Delta\left(  \mathbf{k}\right)  =-V^{-1}\sum_{\mathbf{k}^{\prime}}U\left(
\mathbf{k},\mathbf{k}^{\prime}\right)  K\left(  k^{\prime}\right)
\Delta\left(  \mathbf{k}\right)  , \label{critical temperature}%
\end{equation}
with the understanding that $\Delta\left(  \mathbf{k}\right)  =\Delta
^{s}\left(  \mathbf{k}\right)  $ and $U\left(  \mathbf{k},\mathbf{k}^{\prime
}\right)  =U^{s}\left(  \mathbf{k},\mathbf{k}^{\prime}\right)  $ for singlet
pairing and $\Delta\left(  \mathbf{k}\right)  =\Delta^{t,u}\left(
\mathbf{k}\right)  $ and $U\left(  \mathbf{k},\mathbf{k}^{\prime}\right)
=U^{t}\left(  \mathbf{k},\mathbf{k}^{\prime}\right)  $ for triplet pairing,
where $U^{s}\left(  \mathbf{k},\mathbf{k}^{\prime}\right)  =U_{FF}%
+[U_{ind}\left(  \mathbf{k}-\mathbf{k}^{\prime}\right)  +U_{ind}\left(
\mathbf{k}+\mathbf{k}^{\prime}\right)  ]/2$ and $U^{t}\left(  \mathbf{k}%
,\mathbf{k}^{\prime}\right)  =\left[  U_{ind}\left(  \mathbf{k}-\mathbf{k}%
^{\prime}\right)  -U_{ind}\left(  \mathbf{k}+\mathbf{k}^{\prime}\right)
\right]  /2$ are the singlet and triplet potentials that are even and odd
functions of both $\mathbf{k}$ and $\mathbf{k}^{\prime}$, respectively, and
$K\left(  k\right)  =\tanh\left(  \beta\xi_{k}/2\right)  /\left(  2\xi
_{k}\right)  -1/(2\epsilon_{k})$. \ \ \ In arriving at Eq.
(\ref{critical temperature}), we have applied the standard procedure to
renormalize the contact interaction \cite{bijlsma00} and a similar procedure
(but expressed in terms of vertex functions \cite{baranov02}) to renormalize
the dipolar interaction.

Making decompositions: $\Delta\left(  \mathbf{k}\right)  =\sum_{l}\Delta
_{l}\left(  k\right)  Y_{l}^{0}(\mathbf{\hat{k})}$ and $U\left(
\mathbf{k},\mathbf{k}^{\prime}\right)  =4\pi\sum_{l,l^{\prime},m}%
U_{lm,l^{\prime}m}\left(  k,k^{\prime}\right)  Y_{lm}^{\ast}(\mathbf{\hat{k}%
)}Y_{l^{\prime}m}(\mathbf{\hat{k}}^{\prime})$, in which the azimuthal symmetry
of the interaction (\ref{UIND}) has been explicitly incorporated, we cast
Eq.(\ref{critical temperature}) into the form containing only the radial
coordinate:%
\begin{equation}
\Delta_{l}\left(  k\right)  =-\sum_{l^{\prime}}\int\frac{k^{\prime2}}{2\pi
^{2}}K\left(  k^{\prime}\right)  U_{l0,l^{\prime}0}\left(  k,k^{\prime
}\right)  \Delta_{l^{\prime}}\left(  k^{\prime}\right)  dk^{\prime
},\label{Delta l exact}%
\end{equation}
where $Y_{lm}(\mathbf{\hat{k})}$ are spherical harmonic functions. \ In the
low temperature limit $k_{B}T/\epsilon_{F}\ll 1$, $k^{2}K\left(  k\right)  $ is
small virtually everywhere except around the Fermi momentum $k_{F}$ [$=\left(
3\pi^{2}n_{F}\right)  ^{1/3}$] where it is sharply peaked compared to other
momentum distributions, and the critical temperature can be estimated, to a
good approximation, by the equation for the gap parameter, $\Delta_{l}%
\equiv\Delta_{l}\left(  k_{F}\right)  $, at the Fermi surface
\begin{equation}
\Delta_{l}=N\left(  \epsilon_{F}\right)  \ln\frac{\pi k_{B}T}{8\epsilon
_{F}e^{\gamma-2}}\sum_{l^{\prime}}U_{l,l^{\prime}}\Delta_{l^{\prime}%
}\label{Delta l}%
\end{equation}
where $\gamma=0.577$ is Euler's constant, and $U_{l,l^{\prime}}\equiv
U_{l0,l^{\prime}0}\left(  k_{F},k_{F}\right)  $ is an element of the
interaction matrix $U$: $U_{l,l^{\prime}}^{s}=U_{FF}\delta_{l,0}%
\delta_{l^{\prime},0}+0.5[1+\left(  -1\right)  ^{l^{\prime}}]U_{l,l^{\prime}%
}^{ind}$ and $U_{l,l^{\prime}}^{t}=0.5[1-\left(  -1\right)  ^{l^{\prime}%
}]U_{l,l^{\prime}}^{ind}$, with
\begin{align}
U_{l,l^{\prime}}^{ind} &  =2\pi\int\int\left[  \int U_{ind}(k_{F}%
\mathbf{\hat{k}}-k_{F}\mathbf{\hat{k}}^{\prime})d(\phi_{\mathbf{k}}%
-\phi_{\mathbf{k}^{\prime}})\right]  \times\nonumber\\
&  Y_{l0}\left(  \cos\theta_{\mathbf{k}}\right)  Y_{l^{\prime}0}\left(
\cos\theta_{\mathbf{k}^{\prime}}\right)  d\left(  \cos\theta_{\mathbf{k}%
}\right)  d\left(  \cos\theta_{\mathbf{k}^{\prime}}\right)
.\label{Ull induced}%
\end{align}
In contrast to the dipole-dipole interaction, which couples $\Delta_{l}$ only
to $\Delta_{l\pm2}$, the dipole induced interaction in Eq. (\ref{UIND}) can,
in principle, couple $\Delta_{l}$ to any $\Delta_{(l+2n)\geq0}$ with $n$ being
an integer according to Eq. (\ref{Ull induced}). As expected, the singlet and
triplet pairings are decoupled, containing all the even and odd partial wave
components, respectively.

To solve Eq. (\ref{Delta l}), we move into a primed space in which the
interaction matrix $U^{\prime}=MUM^{\dag}$ is diagonalized via a unitary
transformation $M$ so that $U_{n,n^{\prime}}^{\prime}=-\omega_{n}\left(
U_{BF}^{2}/U_{BB}\right)  \delta_{n,n^{\prime}}$,where $\omega_{n}$ is the
eigenvalue scaled to $-U_{BF}^{2}/U_{BB}$. \ In this primed space, Eqs.
(\ref{Delta l}) are decoupled, leading, immediately, to the critical
temperature
\begin{equation}
T_{n}=\frac{8\epsilon_{F}e^{\gamma-2}}{\pi k_{B}}\exp\left[  -\frac{2}%
{\omega_{n}\left(  \delta\right)  \lambda}\right]  ,
\label{true critical temperature}%
\end{equation}
for the $n$th channel, in which the order parameter $\Delta\left(
\mathbf{k}\right)  \propto\sum_{l}M_{n,l}^{\ast}Y_{l0}(\mathbf{\hat{k})}$
becomes a superposition of different partial waves with different angular
momenta, where
\begin{align}
\delta &  =\xi_{B}k_{F}=k_{F}/\left(  4\sqrt{\pi n_{B}a_{BB}}\right)
,\label{delta}\\
\lambda &  =2N\left(  \epsilon_{F}\right)  \frac{U_{BF}^{2}}{U_{BB}}=\frac
{4}{\pi}\frac{m_{B}m_{F}}{m_{BF}^{2}}\frac{a_{BF}^{2}}{a_{BB}}k_{F}.
\label{lambda}%
\end{align}
As the temperature is lowered, the most favorable superfluid phases to be
realized correspond to those channels with the strongest attractive
interactions (the highest positive $\omega_{n}$).%
\begin{figure}
[ptb]
\begin{center}
\includegraphics[
height=1.6907in,
width=3.3754in
]%
{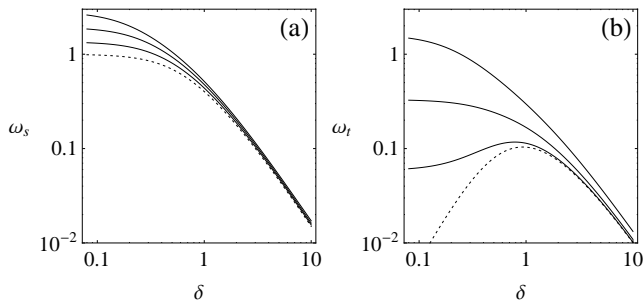}%
\caption{The scaled eigenvalue $\omega_{s}$ for the singlet state (a) and
$\omega_{t}$ for the triplet state (b) as functions of $\delta$ for different
$\varepsilon_{dd}$. \ In both figures, the dotted curves are\ for the nonpolar
case ($\varepsilon_{dd}=0$) and the solid curves from bottom to top are for
(a) $\varepsilon_{dd}=0$.6, 0.8 and 0.9, and (b) $\varepsilon_{dd}=$0.1, 0.5
and 0.9.}%
\label{Fig:EigenFrequencies}%
\end{center}
\end{figure}

Figure \ref{Fig:EigenFrequencies} shows how the strongest (attractive)
interactions in the singlet and triplet channels, $\omega_{s}$ and $\omega
_{t}$, change with $\delta$ for different $\varepsilon_{dd}$ when $U_{FF}=0$.
\ The dotted curves, $\omega_{s}=\left(  2\delta\right)  ^{-2}\ln[1+\left(
2\delta\right)  ^{2}]$ and $\omega_{t}=2\left(  2\delta\right)  ^{-2}\left\{
\ln[1+\left(  2\delta\right)  ^{2}][\left(  2\delta\right)  ^{-2}%
+2^{-1}]-1\right\}  $, represent the corresponding interactions in mixtures
with nonpolar molecules ($\varepsilon_{dd}=0$) \cite{efremov02,bulgac06},
where, as $\delta$ reduces, the triplet interaction begins to decrease to zero
after reaching its peak around $\delta\approx$ $1$ as opposed to the singlet
interaction, which increases monotonically. \ The introduction of a dipolar
condensate to a fermion gas adds to the denominator of Eq. (\ref{UIND}) an
anisotropic term $2\varepsilon_{dd}P_{2}\left(  \cos\theta_{\mathbf{k}%
}\right)  $, which plays an increasingly important role compared to the
isotropic contribution $1+\delta^{2}$ in the small $\delta$ region, around
which significant changes are observed to take place in Fig.
\ref{Fig:EigenFrequencies}. \ The most striking development happens, however,
in the triplet interaction [Fig. \ref{Fig:EigenFrequencies}(b)] which not only
asymptotes to a finite value in the limit $\delta\rightarrow0$, in clear
defiance of Wigner's threshold law, but also increases with $\varepsilon_{dd}%
$\ for a given $\delta$ far more dramatically than the singlet interaction
[Fig. \ref{Fig:EigenFrequencies}(a)]. \ This provides concrete evidence that
the use of a dipolar BEC can indeed significantly enhance scattering in the
triplet channel compared to the singlet channel, which has been the key
motivation behind the present proposal for achieving the triplet superfluid.

To prepare a system with small $\delta$, we must employ a Bose component with
a relatively small (high) healing length (density). \ This often means that
the system separates into a mixed phase with densities $\left(  n_{F1}%
,n_{B1}\right)  $ and a pure Fermi phase with densities $\left(  n_{F2}%
,n_{B2}=0\right)  $, the only phase separation scenario that involves a mixed
phase \cite{viverit00,fodor10}. \ (A complete separation between fermions and
bosons requires much higher densities than considered in the present work.)
\ The mixed phase must share the same chemical and thermodynamical potentials
with the pure phase. \ This consideration leads to
\begin{align*}
U_{BF}n_{B1}+An_{F1}^{2/3}  &  =An_{F2}^{2/3},\\
-U_{BB}n_{B1}^{2}/2-U_{BF}n_{B1}n_{F1}-2An_{F1}^{5/3}/5  &  =-2An_{F2}%
^{5/3}/5,
\end{align*}
from which one finds
\begin{equation}
n_{B1}=An_{F1}^{2/3}\left(  y^{2}-1\right)  /U_{BF}, \label{nB1}%
\end{equation}
where $A=$ $\left(  3\pi^{2}\right)  ^{2/3}\hbar^{2}/\left(  2m_{F}\right)  $
and $y=\left(  n_{F2}/n_{F1}\right)  ^{1/3}$ is the solution to the cubic
equation
\begin{equation}
-15\left(  y+1\right)  ^{2}/\lambda+8y^{3}+16y^{2}+24y+12=0, \label{cubic}%
\end{equation}
(see Ref. \cite{viverit00} for details).\ All previously derived formulas
concerning the critical temperature are directly applicable to the mixed phase
upon substitution of $n_{B}$ with $n_{B1}$ and $n_{F}$ with $n_{F1}$.

\begin{figure}
[ptb]
\begin{center}
\includegraphics[
height=1.8447in,
width=3.3434in
]%
{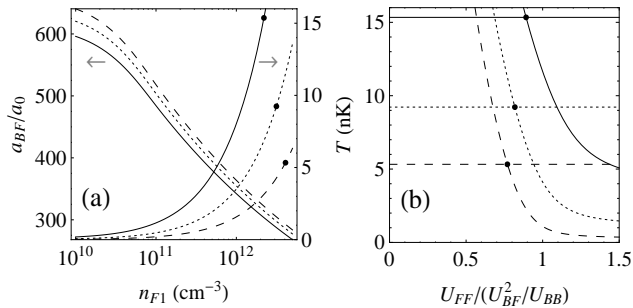}%
\caption{(a) The optimal $T_{t}$ and the required $a_{BF}$ as functions of
$n_{F_{1}}$ for different $\epsilon_{dd}$ under conditions that $m_{F}=6u$,
$m_{B}=127u$ and $a_{BB}=250a_{0}$. (b) illustrates how $T_{s}$ (curves) can
be made degenerate with $T_{t}$ (horizontal lines) by changing $U_{FF}$ while
fixing all the other parameters to those represented by the black dots in (a)
where $n_{B1}$ have reached $5\times10^{14}$ cm$^{-3}$. \ In both figures,
$\epsilon_{dd}=0.9$ for solid curves, $\epsilon_{dd}=0.85$ for dotted curves,
and $\epsilon_{dd}=0.8$ for dashed curves. }%
\label{Fig:temperature}%
\end{center}
\end{figure}

The optimal triplet superfluid temperature $T_{t}$ is always found to occur in
the mixed state. An example in which $m_{F}=6u$, $m_{B}=127u$, and
$a_{BB}=250a_{0}$ (with $u$ the atomic mass and $a_{0}$ the Bohr radius) is
given in Fig. \ \ref{Fig:temperature} (a), which illustrates how $T_{t}$ and
the required $a_{BF}$ change with $n_{F1}$ for different $\varepsilon_{dd}$.
\ In arriving at Fig. \ \ref{Fig:temperature} (a), we used Eq.
(\ref{true critical temperature}) to construct, for a given set of $n_{F1}$
and $\varepsilon_{dd}$, temperature $T_{t}$ as a function of $\lambda$ by
solving for the required $a_{BF}$, $n_{B1}$, and $\delta$ simultaneously from
Eqs. (\ref{delta}) - (\ref{cubic}). \ The optimal $T_{t}$ corresponds to the
peak temperature at some $\lambda=\lambda_{\text{peak}}$. \ In Fig.
\ \ref{Fig:temperature} (a), we find that $\lambda_{\text{peak}}\approx0.6$
and $\delta_{\text{peak}}\approx0.25$ with small variation for different
values of $n_{F1}$ and $\varepsilon_{dd}$. \ As can be seen from the solid
curves ($\varepsilon_{dd}=0.9$), a temperature about 15 nK can be achieved in
a mixed phase (marked with a black dot) with densities ($n_{B1}=5\times
10^{14}$ cm$^{-3}$, $n_{F1}=2.17\times10^{12}$ cm$^{-3}$) and $a_{BF}%
=304a_{0}$. \ \

As to the singlet superfluid temperature $T_{s}$, it depends on the contact
interaction $U_{FF}$ [or $\left(  k_{F}a_{FF}\right)  ^{-1}$], which in our
model is made magnetically tunable via Feshbach resonance. \ Thus, in
principle, the interaction in the singlet channel can be made degenerate to
that in the triplet channel by tuning both the dipolar interaction
($\varepsilon_{dd}$) with an electric field and the s-wave scattering length
($a_{FF}$) with a magnetic field. \ This opens up the possibility of studying
phase competition between singlet and triplet superfluids, a recurring theme
in current studies concerning two-component dipolar Fermi gases. \ Figure
\ref{Fig:temperature} (b) illustrates how $T_{s}$ (curves) can be made to
cross $T_{t}$ (horizontal lines) by changing $U_{FF}$ for different
$\varepsilon_{dd}$. \ In contrast to the pure two-component dipolar Fermi gas
model, where, due to the average of the dipolar interaction over all the
directions being zero, the singlet interaction is always less attractive than
the triplet interaction in the absence of $U_{FF}$, and $U_{FF}$ must be tuned
to the negative side of the Feshbach resonance in order to make $T_{s}$
comparable to $T_{t}$ \cite{shi10}, the singlet interaction in our model is
more attractive than the triplet interaction in the absence of $U_{FF}$ [Fig.
\ref{Fig:EigenFrequencies}], and consequently only when $U_{FF}$ is tuned to
the positive side of the Feshbach resonance, can $T_{s}$ be brought down to
the level of $T_{t}$ [Fig. \ref{Fig:temperature} (b)]. \ An estimate based on
$U_{FF}/\left(  U_{BF}^{2}/U_{BB}\right)  \approx1.08$ at the crossing of the
two solid lines ($\varepsilon_{dd}=0.9$) in Fig. \ref{Fig:temperature} (b)
indicates that $T_{s}=T_{t}\ $when $\left(  k_{F}a_{F}\right)  ^{-1}=2.03$. \ 

In summary, we have investigated the optimal conditions for achieving the
coexistence between singlet and triplet superfluids in a two-component FB
mixture with a dipolar condensate. \ We have found that $T_{s}$ can be made
degenerate to $T_{t}\ $\ at a temperature 10$^{7}$ orders of magnitude higher
than 10$^{-6}$ $nK$ (not shown), the optimal temperature achievable under a
similar set of fixed parameters for a two-component FB mixture with nondipolar
bosons. Just as mixing nonlinear waves has been an important means for
creating coherent sources of laser light, mixing cold atoms is expected to
play an increasingly more important role in creating new quantum gases (or
liquids) in the coming years as the field of ultracold atomic physics
continues to mature. \ The present study reinforces the notion that mixing
fermions with dipolar bosons adds another exciting dimension in the pursuit of
quantum systems with new and novel properties.

H. Y. L. acknowledges the support from the US National Science Foundation and
the US Army Research Office.

\bigskip%

\end{document}